# High Thermal Boundary Conductance across Bonded Heterogeneous GaN-SiC Interfaces

Fengwen Mu,[1,5,a),*] Zhe Cheng,[2,a)] Jingjing Shi,[2] Seongbin Shin,[1] Bin Xu,[3] Junichiro Shiomi,[3] Samuel Graham[2,4,*] Tadatomo Suga[1]

[1] Collaborative Research Center, Meisei University, Hino-shi, Tokyo 191-8506, Japan

[2] George W. Woodruff School of Mechanical Engineering, Georgia Institute of Technology, Atlanta, Georgia 30332, USA

[3] Department of Mechanical Engineering, The University of Tokyo, Bunkyo, Tokyo 113-8656, Japan

[4] School of Materials Science and Engineering, Georgia Institute of Technology, Atlanta, Georgia 30332, USA

[5] Kagami Memorial Research Institute for Materials Science and Technology, Waseda University, Shinjuku, Tokyo 169-0051, Japan

[a)] These authors contribute equally

[*] Corresponding authors: mufengwen123@gmail.com; sgraham@gatech.edu

## Abstract

GaN-based high electron mobility transistors have the potential to be widely used in high-power and high-frequency electronics while their maximum output powers are limited by high channel temperature induced by near-junction Joule-heating, which degrades device performance and reliability. Increasing the thermal boundary conductance (TBC) between GaN and SiC will aid in the heat dissipation of GaN-on-SiC power devices, taking advantage of the high thermal conductivity of the SiC substrate. However, a good understanding of the TBC of this technically important interface is still lacking due to the complicated nature of interfacial heat transport. With the AlN being the typical interfacial layer between GaN and SiC, there are issues concerning the quality of the AlN as well as the defects that are contained in the GaN near this growth interface which can impede heat flow. In this work, a lattice-mismatch-insensitive surface activated bonding method is used to bond GaN directly to SiC and thus eliminating the AlN layer altogether. This allows for the direct integration of high quality GaN layers with SiC to create a high thermal boundary conductance interface. Time-domain thermoreflectance (TDTR) is used to measure the thermal properties of the GaN thermal conductivity and GaN-SiC TBC. The measured GaN thermal conductivity is larger than that of GaN grown by molecular-beam epitaxy (MBE) on SiC, showing the impact of reducing the dislocations in the GaN near the interface. High GaN-SiC TBC is observed for the bonded GaN-SiC interfaces, especially for the annealed interface whose TBC (230 MW/m$^2$-K) is close to the highest values ever reported. Thus, this method provides the benefit of both a high TBC with higher GaN thermal conductivity near the interface to aid in heat dissipation. To understand the structure-thermal property relation, STEM and EELS are used to characterize the interface structure. The results show that, for the as-bonded sample, there exists an amorphous layer near the interface for the as bonded samples. This amorphous layer is crystallized upon annealing, leading to the high TBC found in our work. Our work not only paves the way for thermal transport across bonded interfaces where bonding and local chemistry are tunable, which will enable and stimulate future study of new theory of interfacial thermal transport mechanism, but also impact real-world applications of semiconductor integration and packaging where thermal dissipation always plays an important role.

## Introduction

Gallium nitride (GaN) is an excellent wide bandgap semiconductor for power and RF electronics.[1] GaN-based high electron mobility transistors (HEMTs) have the potential to be widely used in high-power and high-frequency electronics while their maximum output powers are limited by high channel temperature induced by near-junction Joule-heating, which degrades device performance and reliability.[2,3] Proper thermal management are the key to these devices for stable performance and long lifetime. Due to the high thermal conductivity and relatively small lattice mismatch with GaN, SiC is usually used as the substrates for high-power applications. However, the thermal boundary conductance (TBC) between GaN and SiC limits the effectiveness for heat dissipation from GaN to SiC.[4] For this technically important interface, a number of experimental and simulation studies have been reported to understand thermal transport across the GaN-SiC interfaces.[4-12] The calculated TBC of GaN-SiC interface is close to 500 MW/m$^2$-K by molecular dynamics (MD) simulations, twice as the experimentally measured GaN-SiC TBC.[4-9] Only a first-principle calculation matches with experimental values,[12] but acoustic mismatch model (AMM) and diffusive mismatch model (DMM) are used to calculate transmission in their work, which did not include the contribution of inelastic scatterings and cannot address the problem of local non-equilibrium phonon transport near the interface.[13] Therefore, a unified understanding of thermal transport across GaN-SiC interface is still lacking because of the complicated nature of interfacial heat transport. The complications arise due to interfacial layers such as AlN that are used in between the GaN and SiC and the resulting quality of the GaN (e.g., dislocation density) that exists when GaN is grown onto the AlN. Both of these will add to the thermal resistance of the device structure near the interface. In general, what impact TBC are not only common factors such as temperature and phonon dispersion relations of the involved materials, but also interfacial bonding and local chemistry near the interfaces.[14,15]

The GaN-SiC interfaces reported in the literature are generally grown by MBE or metalorganic chemical vapor deposition (MOCVD) with a layer of AlN transition layer, which is necessary to grow high-quality GaN due to the lattice mismatch between GaN and SiC. This AlN layer could serve as a phonon bridge to enhance TBC between GaN and SiC with a proper thickness because of the large phonon density of states mismatch of GaN and SiC but it has not been verified for the GaN-SiC interface.[16] Besides growing GaN on SiC, room-temperature surface activated bonding (SAB) is an important technique which has the potential to be widely used in the heterogeneous integration of semiconductor materials and in microelectronics packaging.[17,18] Compared with MBE, SAB is insensitive to lattice mismatch and can be performed at room temperature and at wafer-scale which results in small thermal stress. The bonded interfaces can have different interfacial bonds and local chemistry from directly-grown interfaces, and provide novel interfaces which cannot be grown through other techniques. This will enable and stimulate future studies of interfacial thermal transport mechanisms with such high degree of control over heterogeneous interfaces. Additionally, from an applied point of view, the thermal transport properties across these bonded interfaces are of great significance for real-world applications of electronics integration and packaging where heat dissipation is always an important issue.[3]

In this work, template GaN is bonded with a 4H-SiC substrate followed by thinning it to several hundred nanometers for interfacial thermal transport characterization. TDTR is used to measure the thermal properties of GaN layer and GaN-SiC TBC for samples with and without post-annealing. The thermal conductivity of the GaN layer is compared with that of MBE-grown GaN. To understand the structure-thermal property relation, high-resolution scanning transmission electron microscopy (HR-STEM) and electron energy loss spectroscopy (EELS) are used to study the interface structure and local chemistry distribution.

**Results and Discussion**

To fabricate our samples, a 2-μm-thick GaN template is bonded with a 4H-SiC substrate by SAB before applying a laser lift-off process. To get good sensitivity of the GaN-SiC TBC, the GaN layer is thinned down before depositing a ~70 nm Al layer as TDTR transducer. More details about the bonding process can be found in the experimental sections. Here, three samples are prepared: a bare SiC substrate, a GaN-SiC as-bonded sample, and a GaN-SiC bonded sample with 1000 ºC post-annealing. The thermal conductivity of the SiC substrate is measured and the results are used as input in the data fitting of the other two samples. The picosecond acoustic technique is used to measure the local Al and GaN thicknesses. The measured thicknesses are confirmed with TEM and the results show excellent agreements. The thermal properties of the samples in this work are measured by TDTR, a pump-probe method widely used to measure the thermal properties of both nanostructured and bulk materials.[19,20] A modulated pump beam heats the sample surface periodically and a delayed probe beam detects the surface temperature variation via thermoreflectance. The signal picked up by a photodiode and a lock-in amplifier is fitted with an analytical heat conduction solution to infer the unknown parameters. More details about TDTR can be found in the experimental section.

Multiple spots are measured on the two bonded samples. Because the GaN thickness is not uniform after thinning, we obtain thickness dependence of GaN thermal conductivity and GaN-SiC TBC, as shown in Figure 1. As shown in Figure 1 (a), the measured GaN thermal conductivity is slightly larger than that of the MBE-grown GaN. The error bars of normal TDTR measurements are ±10%. The MBE-grown GaN is directly grown on SiC so the lattice mismatch leads to a relatively large amount of dislocations and defects in the thin GaN film near the interface. Phonons scattering with these structural imperfections lead to reduced mean free paths and correspondingly reduced thermal conductivity. In this work, commercially available template GaN is used in our bonding which has less structural imperfections. Here, SAB provides a solution to avoid the integration of low-quality GaN into sample structures which cannot be avoided when direct growth with MOCVD is used. The SAB bonding method eliminates the growth of transition layers and defective GaN near the growth interface where high concentrations of defects and relatively low thermal conductivity are expected. Also, the GaN thermal conductivity increases with GaN film thickness due to the phonon-boundary scattering. 80% of the thermal conductivity of bulk GaN is contributed by phonons with mean free paths from 100 nm to 3 μm.[21] Phonons with mean free path longer than the film thickness scatter with the film boundary. The film thickness limits the phonon mean free path and limits the thermal conductivity. For the ~600 nm and ~300 nm thick GaN films, their thermal conductivity are reduced to half and one third of the bulk value, respectively.

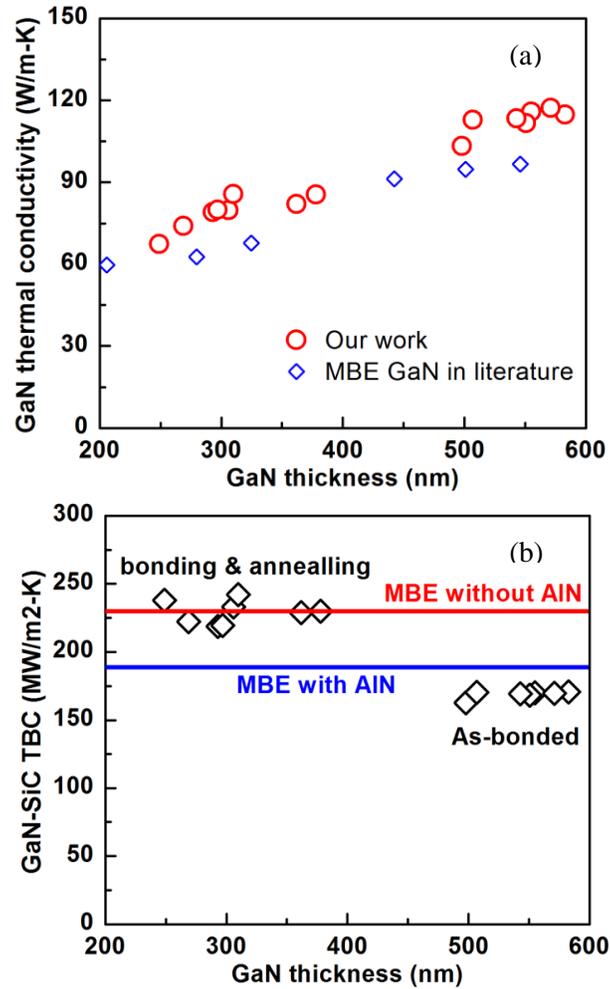

Figure 1. (a) thickness-dependent thermal conductivity of the GaN thin films. The thermal conductivity of MBE-grown GaN is from literature.[22] (b) thickness dependence of GaN-SiC TBC for annealed and as-bonded samples. The TBC of MBE-grown GaN on SiC with and without a AlN transition layer are literature.[4,9]

Figure 1 (b) shows the measured TBC of both the as-bonded and annealed GaN-SiC interfaces. The measured TBC values do not change with GaN thickness as expected. The TBC of the as-bonded GaN-SiC interface is slightly lower than that of the GaN grown on SiC with an AlN transition layer. But after annealing at 1000 °C for 10 mins, the TBC increases to about 230

MW/m$^2$-K, one of the highest reported values measured by experiments.[4] This value is one of the highest GaN-SiC TBC reported ever, as summarized in Table 1. It is notable that we did not see large variation for the TBC values in the cm-scale samples. We randomly measured 15 spots on the as-bonded and annealed samples. The variation of the measured TBC is within 10%, which is within the error bar of TDTR measurements. Since the small original thermal stress of the sample bonded at room temperature as well as the small thermal expansion coefficient mismatch between GaN and SiC, the TBC of the sample still shows uniform high values after experiencing high temperature annealing up to 1000 ºC. This facilitates real-world applications, for example, homoepitaxy-growth of GaN on GaN-SiC bonded structure or device fabrications because epitaxy-growth and electrode formation are high temperature processes. Thus, these data show that we can obtain very high TBC values along with higher thermal conductivity GaN in the vicinity of the interface.

Table 1. Summary of experimentally measured GaN-SiC TBC in the literature and this work

| Literature | AlN layer | Integration method | GaN-SiC TBC | Thermal measurements |
|---|---|---|---|---|
| Ref. 1[4] | No | MBE | 230 MW/m$^2$-K | FDTR |
| Ref. 2[9] | Yes | MBE | 189 MW/m$^2$-K | TDTR |
| Ref. 3[23] | Yes | MOCVD | 200 MW/m$^2$-K | TDTR |
| Ref. 4[11] | Yes | MOCVD | 20-67 MW/m$^2$-K | Raman |
| Ref. 5[10] | Yes | MOCVD | 30 MW/m$^2$-K | Raman |
| This work | No | SAB | 169 MW/m$^2$-K | TDTR |
| This work | No | SAB with annealing | 229 MW/m$^2$-K | TDTR |

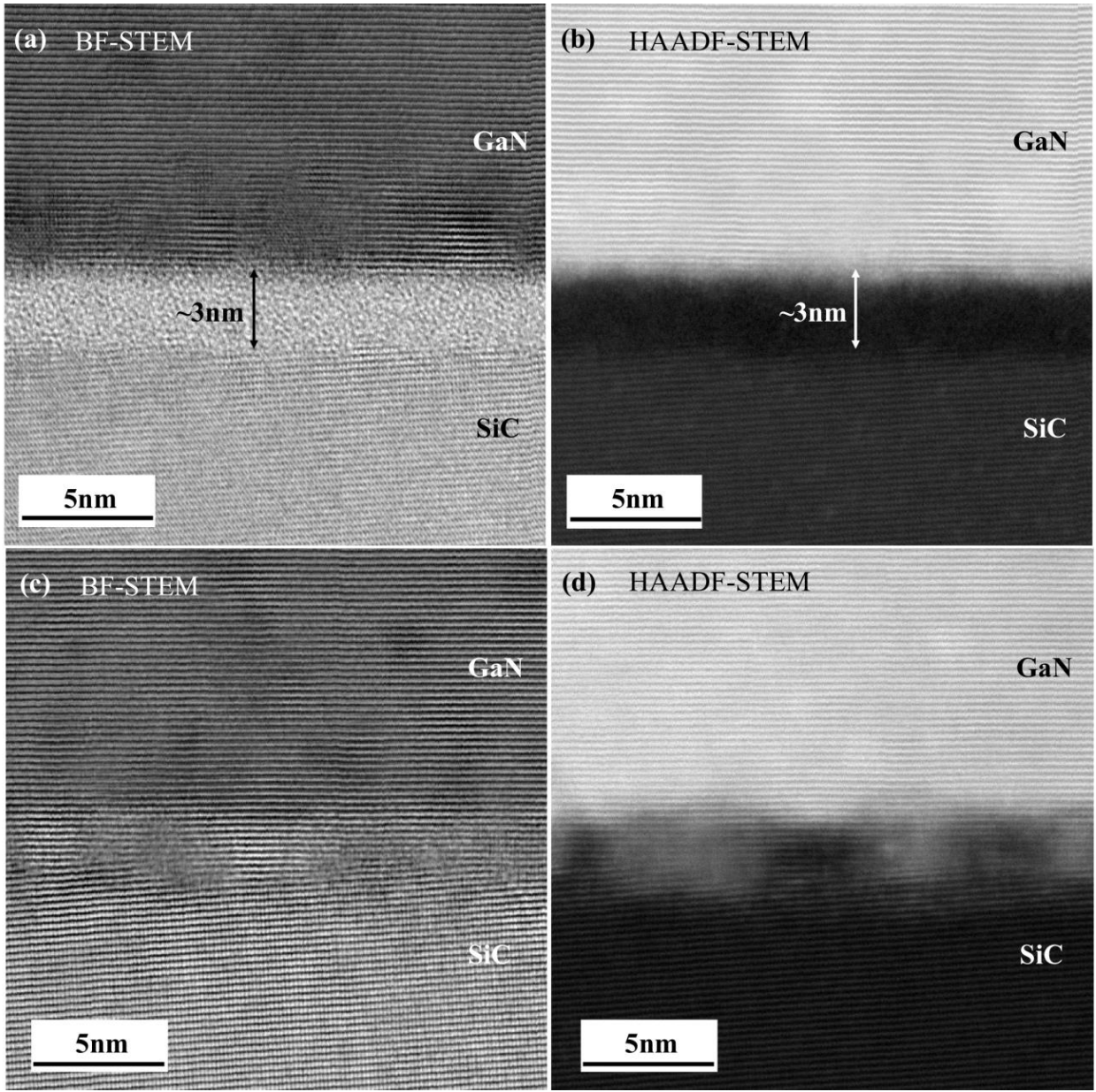

Figure 2. Cross-section HR-STEM images of GaN-SiC interface bonded at room temperature and after annealing at 1000 °C: (a)(c) bright-field (BF) image and (b)(d) high-angle annular dark-field (HAADF) image.

To further understand the thermal property-structure relation, HR-STEM and EELS are used to study the GaN-SiC interfaces. The structure of the as-bonded GaN-SiC interfaces are shown in

Figure 2 (a). There exists a ~3 nm thick amorphous layer at the as-bonded interface, resulting from the ion beam bombardment during surface activation in the bonding process. The amorphous layer is mainly amorphous SiC, where amorphous GaN is hardly recognized. According to the blurred interface between amorphous GaN and amorphous SiC in the HAADF image of the as-bonded interface, as shown in Figure 2 (b), the interfacial diffusion seems already happen even at room temperature. As confirmed by the high resolution EELS analysis in Figure 3(a), the amorphous SiC contains a Si-enriched layer, which is formed by the intentionally-designed Si-containing Ar ion beam. The Si-enrichment of the amorphous SiC is assumed to be helpful for the interfacial diffusion at room temperature, which is consistent with previous bonded interfaces of GaN-Si and GaN-SiC.[24,25] Both of the interfacial amorphous layer and interfacial mixing caused by diffusion may contribute the high TBC of the as-bonded interface.[26] Besides, a small amount of Ar (a couple of percent in atom composition) is only observed in the amorphous SiC layer, which derives from the Ar ion beam in the bonding process.[24] Some simulations show that the implanted Ar in the GaN side is possible to be ejected out.[27] Similar phenomena are also observed in bonded GaN-Si interface.[25] It is still an open question that how these trapped Ar atoms affect thermal transport across interface, especially with an amorphous interface layer where phonon gas model does not hold. Recent MD modeling results show that there exists an interface mode at heterogeneous Si-Ge interfaces.[28-30] The interface mode interacts with phonons in both sides of the interface, redistributes phonon modes and boosts the contribution of inelastic transport to TBC.[31]

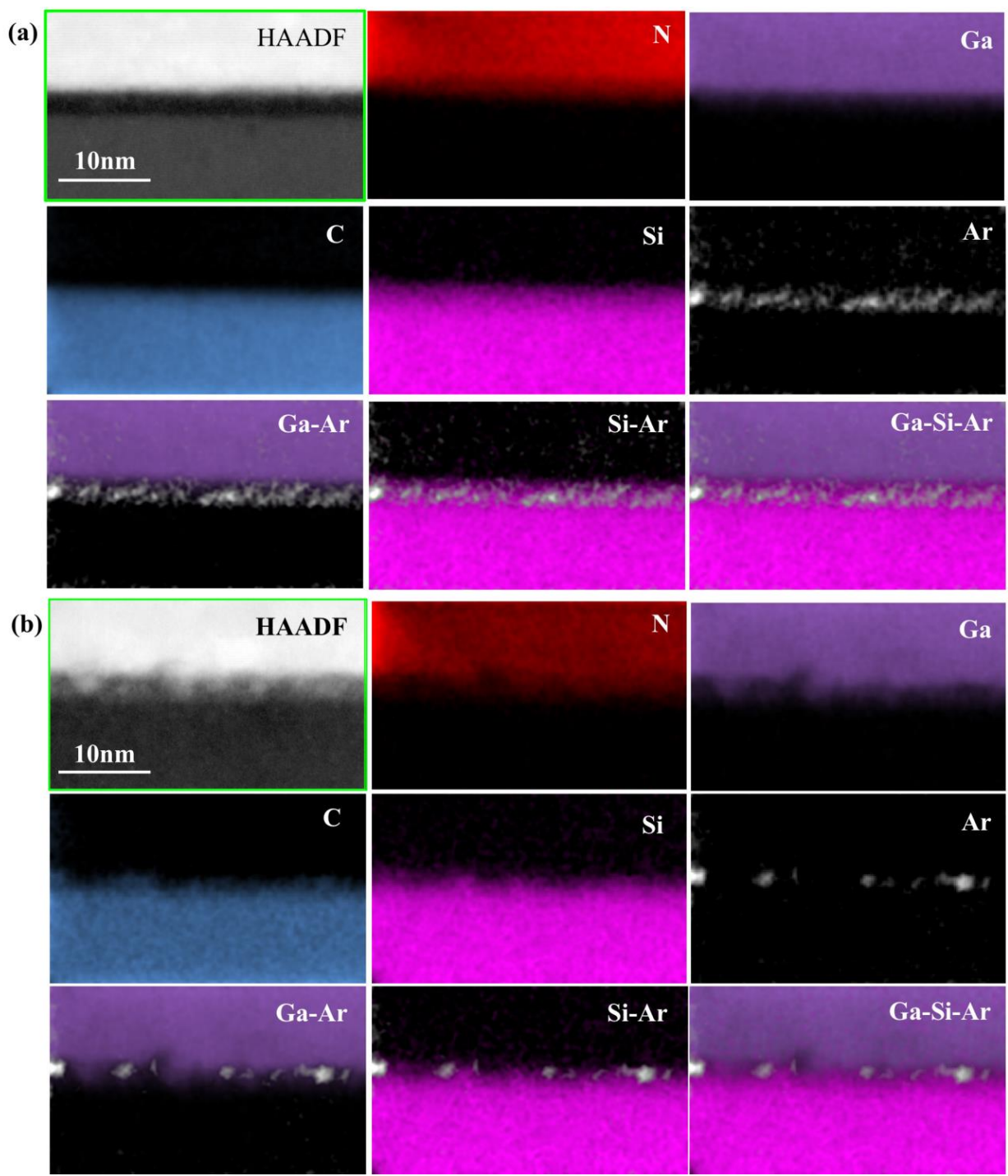

Figure 3. EELS mapping of GaN-SiC bonded interfaces: (a) at room temperature and (b) after annealing at 1000 °C. The N, Ga, C, Si, and Ar maps are highlighted in red, purple, blue, pink, and white, respectively.

As a comparison, the TEM structure and EELS images of the annealed interface are shown in Figure 2 (c-d) and Figure 3 (b), respectively. The amorphous layer crystallizes during annealing and almost disappears in the TEM images. But some local inhomogeneous structure and dislocations show up near the interface. This is possibly due to the non-uniform interfacial diffusion during annealing, as shown in Figure 2(d) and Figure 3(b). The trapped Ar atoms also redistribute and form islands at the interface but still in SiC side, as shown in the EELS images. An amorphous layer usually has very low thermal conductivity. After annealing, the measured TBC increases by 36%, to almost the same value as the TBC of directly-grown GaN on SiC. Here, we attribute this increase in TBC to the disappearance of the amorphous layer and the redistribution of the Ar atoms even though some modeling results show that amorphous interface increases TBC.[20,30] We also calculate the GaN-SiC TBC by a Landauer approach with DMM and the calculated TBC is about 260 MW/m$^2$-K, close to the literature value.[12] If assuming the phonon transmission is unity (all the phonons in the GaN side transmit across the interface to the SiC), we obtain the radiation limit as 367 MW/m$^2$-K. This gives us an estimate of the upper limit of GaN-SiC TBC if inelastic scattering at the interface is not significant.

## Conclusions

In this work, we bonded a GaN layer with a SiC substrate by a room temperature bonding method which brings bulk-quality GaN directly to the GaN-SiC interface, which facilitates thermal dissipation for devices. Moreover, a high GaN-SiC TBC is observed for the bonded GaN-SiC interface, especially for the annealed interface whose TBC (230 MW/m$^2$-K) is close to the highest reported values in the literature. STEM and EELS results show that, for the as-bonded sample, there exists an interfacial amorphous layer and a small amount of trapped Ar atoms at the interface while the amorphous layer disappears and the Ar atoms redistribute after post-annealing. We attribute the TBC increase after annealing to the disappearance of the amorphous layer and the redistribution of the trapped Ar atoms. Our work not only paves the way for thermal transport across bonded interfaces where bonding and local chemistry are tunable, which will enable the study of interfacial thermal transport mechanisms, but also impact real-world applications of semiconductor integration and packaging where thermal dissipation always plays an important role.

**Experimental Sections**

**Sample Preparation**: a Ga-face 2-μm GaN layer grown on a sapphire substrate was bonded to a Si-face, 3-inch, 4º off, 4H-SiC wafer by a SAB machine at room temperature. The root-mean-square (RMS) surface roughness of the GaN and SiC surface is ~0.4 nm and ~0.3 nm, respectively. After the GaN and SiC surfaces are activated by Ar ion beam and intentionally-designed Si-containing Ar ion beam, the two wafers are bonded at room temperature with a pressure of 5MPa for 300s. The sapphire substrate was removed by a laser-lift-off process. To obtain good TDTR sensitivity for the buried GaN-SiC interface, the GaN layer was thinned to 300~600 nm thick by polishing. The bonded wafer was diced into chips. To study the effect of thermal annealing on TBC, one chip was annealed at 1000 ºC for 10 min in a flowing $N_2$ gas. After that, a ~70nm Al layer was deposited on the samples by sputtering as TDTR transducer. Another bonded wafer obtained at the same condition was used for the evaluation of bonding energy by the "crack-opening" method.[32]

**TDTR**: TDTR measurements are performed on the as-bonded and annealed samples to extract the GaN thermal conductivity and GaN-SiC TBC with a 10X objective and a modulation frequency of 3.6 MHz. The sensitivity of TDTR to each unknown parameter is shown in the Supplementary Materials. The sensitivity of GaN-SiC TBC is very high and good for accurate measurements.[33] The fitting of the experimental data and the analytical curve is excellent. The fitting parameters are also included in Supplementary Materials.

**HRTEM-EELS**: Cross-section focus ion beam (FIB) TEM samples were prepared with a FEI Helios dual beam FIB/SEM system. The interface structures were characterized by an aberration-corrected STEM (Hitachi HD2700) and the interface composition was measured by EELS (Gatan Enfinium) with a step size of 0.2 nm. Since the Si-face of SiC has 4º off-axis towards <11-20>, to avoid interface overlap, the observation in this study is along <1-100> axis for both of SiC and GaN.

## Acknowledgements

F.M., S.S., and T.S. would like to acknowledge the financial support from R&D for Expansion of Radio Wave Resources, organized by the Ministry of Internal Affairs and Communications, Japan and JSPS KAKENHI Grant Number 19K15298. Z.C., J.S., and S. G. would like to acknowledge the financial support from ONR MURI Grant No. N00014-18-1-2429.

## Completing Financial Interest

The authors claim no completing financial interest.

# High Thermal Boundary Conductance across Bonded Heterogeneous GaN-SiC Interfaces

Fengwen Mu,[a),*] Zhe Cheng,[a)] Jingjing Shi, Seongbin Shin, Bin Xu, Junichiro Shiomi, Samuel Graham[*] Tadatomo Suga

[a)] These authors contribute equally

[*] Corresponding authors: mufengwen123@gmail.com; sgraham@gatech.edu

The employed surface activated bonding (SAB) machine consists of a load-lock chamber and a processing-bonding chamber. The background vacuum of the processing-bonding chamber is kept as $5\times10^{-6}$ Pa. Two kinds of Ar ion beam sources (Ar ion beam and Si-containing Ar ion beam) were setting in the processing-bonding chamber for surface activation. The Ar ion beam is for GaN surface and the Si-containing Ar ion beam was intentionally designed for SiC to suppress its Si preferential sputtering during surface activation.[1] Both of them have a power of 1.0 kV and 100 mA. After the surface activation for both GaN and SiC, two wafers were bonded at room temperature by contact with a pressure of 5MPa at room temperature for 300s. The sapphire substrate of GaN template was removed from the bonded wafer by laser-lift-off method using a laser with a wavelength of 248 nm, followed by a polishing process for thinning. The GaN layer was thinned to 300~600 nm for an accurate TDTR measurement. Then, the bonded wafer was diced into chips for different analyses. To investigate the effect of annealing, part of the samples were annealed at 1273 K for 10 min in flowing $N_2$ gas. An ~70nm Al transducer layer was deposited by ion beam sputtering on all of the samples for TDTR measurements. Another bonded wafer obtained at the same condition was used for the evaluation of bonding energy by the "crack-opening" method.[2] The bonding energy ($\gamma$), which is the fracture energy of bonding interface, was evaluated by the "crack-opening" method and calculated by the following equation [1].

$$\gamma = \frac{3t_b^2 E_1 t_{w1}^3 E_2 t_{w2}^3}{16 L^4 \left(E_1 t_{w1}^3 + E_2 t_{w2}^3\right)} \qquad \text{(Equation S1)}$$

where $E_1$ and $E_2$ are the Young's moduli of SiC (530 GPa) and sapphire (345 GPa), $t_{w1}$=0.355 mm and $t_{w2}$=0.432 mm are the thickness of two wafers, $t_b$=0.1 mm is the thickness of the blade, and L is the crack length. In this evaluation, the bonded wafer is simplified to be a bonding between sapphire and SiC, since the thickness of GaN layer is neglected compared to that of sapphire substrate. This measurement was carried out at room temperature in air at a relative humidity (RH) of ~39.6%.

Figure S1(a) shows an image of the bonded wafer. Most area of the two wafers was well bonded except a few bonding voids resulting from particles and some edge parts. After a laser lift-off process, the sapphire substrate can be removed and the GaN layer was transferred onto a SiC substrate, as shown in Figure S1(b). The bonding energy of ~1.3±0.3 J/m$^2$ is achieved. The bonding interface can withstand both the thinning and the cutting processes.

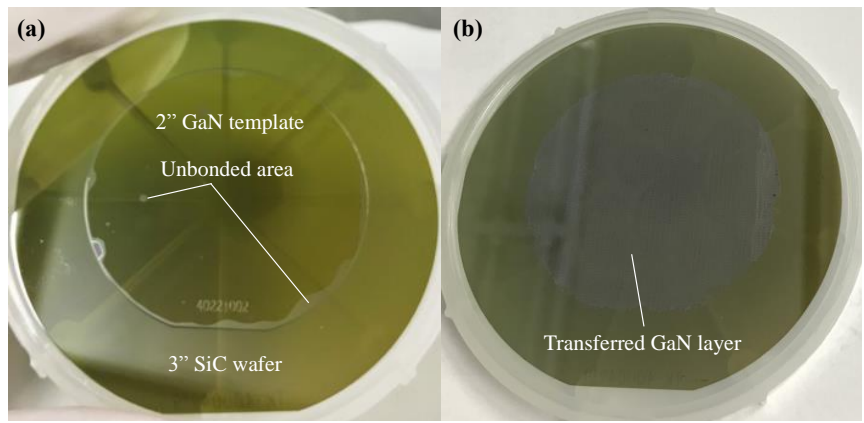

Figure S1. (a) Wafer bonded at room temperature by SAB method and (b) Bonded wafer after a laser lift-off process.

Figure S2 shows the echoes from the picosecond acoustic technique. We measured the Al and GaN thicknesses according to these echoes. The echoes indicate the time of strain waves travel from sample surface to a certain interface and bounce back. If we know the longitudinal acoustic phonon group velocity, then we can obtain the film thickness. This technique has been widely used in almost all the TDTR labs for decades. Excellent agreements have been achieved with thicknesses measured by other techniques such as atomic force microscopy (AFM) and TEM in the literature and this work.

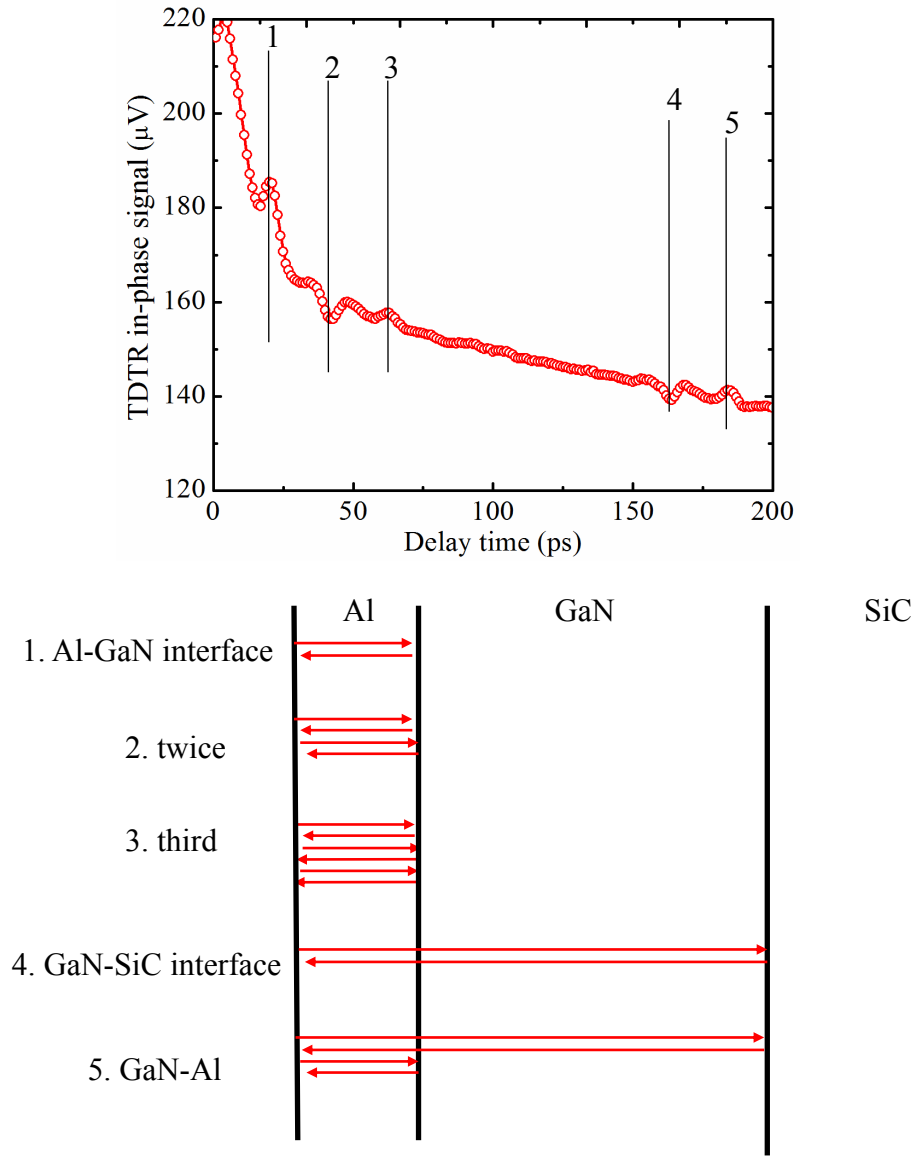

Figure S2. The upper image shows the echoes from the picosecond acoustic technique and the lower image shows what these echoes represent.

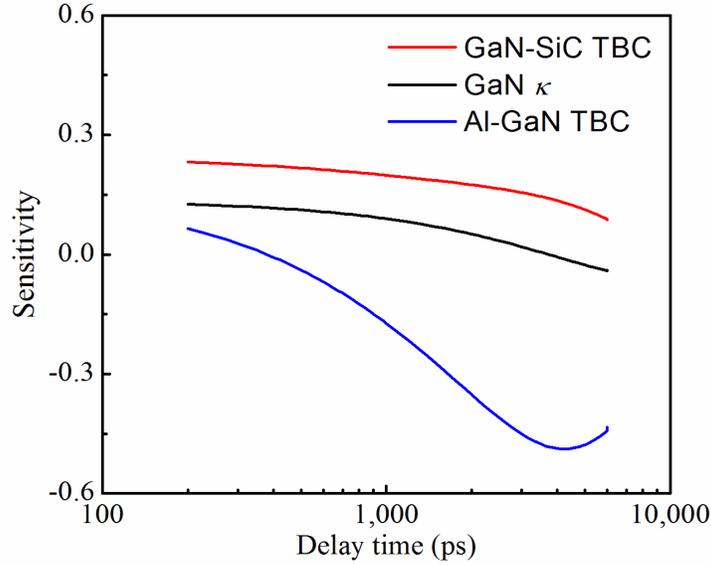

Figure S3. TDTR sensitivity of each fitting parameter. The sensitivity of GaN-SiC TBC (the red line) is very high.

The definition of TDTR sensitivity is

$$S_i = \frac{\partial \ln(-V_{in}/V_{out})}{\partial \ln(p_i)} \quad \text{(Equation S2)}$$

Where $S_i$ is the sensitivity to parameter $i$, $-V_{in}/V_{out}$ is the TDTR signal, $p_i$ is the value of parameter $i$. Figure S3 shows the TDTR sensitivity of each fitting parameter. The sensitivity of GaN-SiC TBC is very high, which is good for accurate measurements.

The thermal conductivity, density, and heat capacity of Al is 150 W/m-K, 2700 kg/m$^3$, and 900 J/kg-K.[1] The thermal conductivity of Al is calculated from electrical conductivity measurements and applying Wiedemann-Franz law. The density and heat capacity of GaN is 6150 kg/m$^3$ and 430 J/kg-K.[2] The thermal conductivity, density, and heat capacity of SiC is 331 W/m-K, 3210 kg/m$^3$, and 690 J/kg-K.[2] The thermal conductivity of the SiC substrate is measured by TDTR in this work.

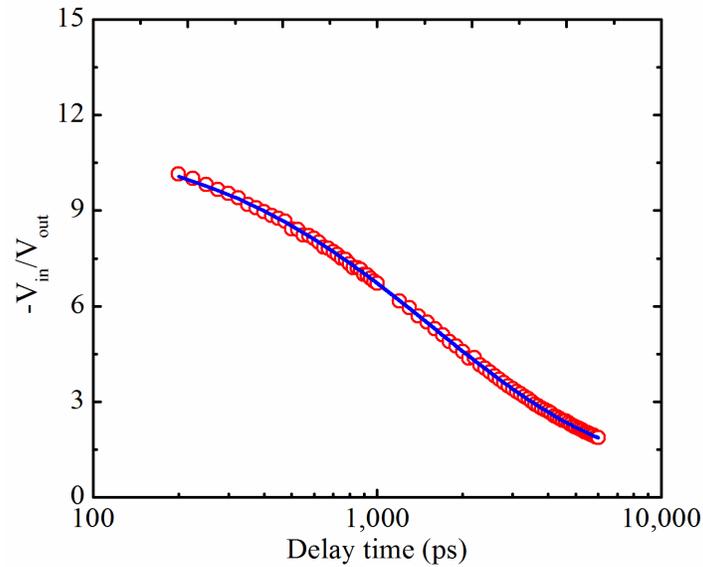

Figure S4. TDTR data fitting. The red circles are experimental data while the blue line is the analytical fitting curve. We can see very good agreement between the experimental data and the theoretical curve.